\documentstyle[revtex]{aps}
\begin{document}
\begin{title}
Enhancement of the decay rate of nonequilibrium carrier
distributions due to scattering-in processes\cite{NIST}
\end{title}
\author{B.A. Sanborn\cite{FEL}}
\begin{instit}
Semiconductor Electronics Division
National Institute of Standards and Technology, Gaithersburg, MD 20899
\end{instit}
\moreauthors{Ben Yu-Kuang Hu and S. Das Sarma}
\begin{instit}
Department of Physics, University of Maryland, College Park, MD 20742-4211
\end{instit}
\vspace{7mm}
\begin{abstract}
We show that, for some semiconductor devices and physical experiments,
 processes which scatter electrons {\em into} a state
$|{\bf k}\rangle$ can contribute strongly to the decay  of
a nonequilibrium electron occupation of $|{\bf k}\rangle$.
For electrons,
the decay rate $\gamma({\bf k})$ is given by the sum of the total
scattering-out {\em and scattering-in} rates of state $|{\bf k}\rangle$.
The scattering-in term, which is often neglected in calculations,
increases $\gamma({\bf k})$ of low energy electrons injected into
semidegenerate systems, which includes many doped semiconductor
structures at nonzero temperatures, particularly those
of reduced dimensions.
\end{abstract}

\vspace{7mm}
Inelastic scattering of electrons in doped semiconductor systems
is of interest because of its technological importance in device
performance and because it provides information on the
fundamental carrier interactions.  For example, it is critical in
determining whether substantial ballistic transport of electrons is
possible through devices of small dimension\cite{YOK,BAL}
and whether coherent resonant tunneling conduction is possible in double
barrier and superlattice devices.\cite{TUN}
It is also used for interpretation of hot
electron\cite{LEV}$^{,}$\cite{HE}
and ultrafast optical\cite{KNO} spectroscopy experiments.

In these situations, a {\em nonequilibrium} distribution
of carriers is injected into
a system in equilibrium, and it is the {\em decay}
of this nonequilibrium distribution, either temporally or spatially,
that determines the measured
experimental quantities ({\em e.g.}, the differential transmission
in an ultrafast optical spectroscopy experiment
or the current in a ballistic transport transistor).
It has generally been assumed that
the decay rate of a nonequilibrium distribution which is  peaked around
momentum ${\bf k}$
will be given completely by the total scattering-out rate
of state $|{\bf k}\rangle$.  In this letter, we point out that, while
this is a good approximation for most experimental situations at very
low temperatures or in metals, it is {\sl not} true in general.
When electrons are injected into states that have a significant
thermal electron occupation, the decay rate is
significantly {\sl larger} than the scattering-out rate.
The decay of the nonequilibrium
distribution function (for fermions) is actually given by the
{\em sum} of the total scattering-out {\em and total scattering-in} rates.

There are three important energy scales in the problem:
the chemical potential $\mu$ and temperature $T$
of the equilibrium system, and the electron
injection energy, $E$.   We show that in situations where
$E-\mu$ is much larger than
$k_BT$ (which includes most experimental situations in
metals, as well as the low-temperature and completely
nondegenerate regimes in doped semiconductors),
the scattering-in rate is negligible and
the nonequilibrium decay rate is well-approximated by the
scattering-out rate alone.
However, when $E-\mu$ is less than or on the order of several $k_BT$,
the scattering-in rate becomes significant and {\sl must} be included
in determining the decay rate of an injected nonequilibrium
distribution.  The often-ignored scattering-in rate could substantially
affect the interpretation of experimental results, especially in
lower dimensional semiconductor nanostructures.

We derive this result using the Boltzmann equation.
Collision processes produce
the time-evolution of the distribution function $f({\bf k},t)$ given by
\begin{eqnarray}
{\partial f({\bf k},t)\over\partial t} =
-\int \frac{d^{d}q}{(2\pi)^{d}} \left\{
P({\bf q},\omega_{{\bf k},{\bf k}-{\bf q}})f({\bf k},t)
[1-f({\bf k}-{\bf q},t)]
\right.  \nonumber \\
-\left. P({\bf q},\omega_{{\bf k}-{\bf q},{\bf k}})f({\bf k}-{\bf q},t)
[1-f({\bf k},t)]
\right\},\label{eq1}
\end{eqnarray}
where $d$ is the system dimensionality,
$P({\bf q},\omega)$ gives the probability per unit time
for a carrier to scatter with change of
momentum ${\bf q}$ and energy $\hbar\omega$,
and $\omega_{{\bf k},{\bf k'}} = \{E({\bf k})-E({\bf k}')\}/\hbar$.

In most of the experiments described above,
the nonequilibrium part of the distribution function
is monoenergetic and hence
sharply peaked in ${\bf k}$-space.
Let's assume a
perturbation $\delta\!f({\bf k},t)$ from
the equilibrium distribution $f_0({\bf k})$ that is not necessarily
small but which is sharply peaked around
${\bf k}={\bf k}_0$, so that $\delta\!f({\bf k},t)$
can be approximated by a $\delta$-function:
\begin{eqnarray}
f({\bf k},t) &=& f_0({\bf k}) + \delta\!f({\bf k},t)\label{eq2}\\
\delta\!f({\bf k},t) &=& g({\bf k}_0,t)\,\delta({\bf k}-{\bf k}_0).
\end{eqnarray}
Substituting Eqs.\ (\ref{eq2}, 3) into Eq.\ (\ref{eq1})
gives the decay rate $\gamma$ of the non-equilibrium distribution
$\delta\! f$.
Noting that the terms not involving $\delta\!f$ cancel
(because there is no net change in $f_0$
due to scattering), we obtain
\begin{eqnarray}
\frac{d g({\bf k}_{0},t)}{dt} &=&  -\gamma({\bf k}_{0})\,
g({\bf k}_{0},t), \label{eq4b} \\
\gamma({\bf k}_{0}) &\equiv& \gamma_{\rm out}({\bf k}_0) +
\gamma_{\rm in}({\bf k}_0),\label{eq4a} \\
\gamma_{\rm out}({\bf k}) &=& \int
\frac{d^{d}q}{(2\pi)^d}  P({\bf q},
\omega_{{\bf k,k-q}})[1-f_0({\bf k}-{\bf q})],\\
\gamma_{\rm in}({\bf k}) &=& \int \frac{d^{d}q}{(2\pi)^d}
P({\bf q},\omega_{{\bf k-q,k}})f_0({\bf k}-{\bf q}).
\end{eqnarray}
Here, $\gamma_{\rm out}({\bf k})$
can be identified as the total equilibrium electron
scattering-out rate from an occupied state ${\bf k}$.
Similarly, $\gamma_{\rm in}({\bf k})$ is the
total equilibrium rate for electron scattering into an unoccupied state
$|{\bf k}\rangle$ (or, equivalently, the equilibrium
conduction band {\em hole}
scattering-out rate from $|{\bf k}\rangle$).\cite{MANYBODY}
The origin of the familiar $\gamma_{\rm out}({\bf k})$ term
is obvious. The $\gamma_{\rm in}({\bf k})$
term is due to the increased nonequilibrium occupation of $|{\bf k}\rangle$
by the injected electron, which blocks by Pauli exclusion those
processes that scatter electrons into $|{\bf k} \rangle$.
The electron scattering-in
rate for the state is thereby reduced below the equilibrium rate,
causing the $\delta\! f({\bf k},t)$ to decay {\em faster} than
$\gamma_{\rm out}({\bf k})$ alone would indicate.

Though the existence of the scattering-in term $\gamma_{\rm in}({\bf k})$
has been recognized for many years,\cite{KAD}
it is often neglected in calculations
of $\gamma({\bf k})$.  We now show that this is justified
only for nondegenerate systems or for injection energies
large compared to the Fermi surface thermal width
above a Fermi sea ({\em e.g.},
electron energy loss experiments in metals).
The condition that there is no net change in the distribution function
at equilibrium implies that the left hand side of Eq.\ (\ref{eq1}) is zero
when $f({\bf k},t)=f_{0}({\bf k})$,
yielding
\begin{equation}
\gamma_{\rm out}({\bf k})f_0({\bf k}) = \gamma_{\rm in}({\bf k})
[1-f_0({\bf k})]\label{eq5}
\end{equation}
which implies, from Eq.\ (\ref{eq4a}),
\begin{equation}
\gamma({\bf k}) = {\gamma_{\rm out}({\bf k}) \over 1 - f_0({\bf k})}.
\end{equation}
That is, the decay rate $\gamma({\bf k})$ {\em is enhanced by the factor}
$[1-f_0({\bf k})]^{-1}$ {\em over the scattering-out
rate} $\gamma_{\rm out}({\bf k})$.
Thus, the $\gamma_{\rm in}({\bf k})$
contribution to the decay of a single-electron excitation of energy $E({\bf
k})$
is significant only if $f_0({\bf k})$ is not small compared to unity.
This justifies the common practice of neglecting
$\gamma_{\rm in}({\bf k})$
in the classical nondegenerate limit\cite{CLA}
 or when the excitation energy is much
larger than the thermal width above the equilibrium system
chemical potential.\cite{YOK,LEV,ZER}
Also, while the enhancement in $\gamma$ is very large
for $E({\bf k})$ below $\mu$,
it is very difficult to perform experiments involving injection of electrons
below $\mu$.
However, for low excitation energies above $\mu$ in the semidegenerate
or degenerate regimes,
$\gamma_{\rm in}({\bf k})$ can be significant,
as we show below.

First, let us make explicit the criterion for the validity of nondegenerate
statistics, where the enhancement factor is
negligible.  The degree of degeneracy of an electron gas
may be characterized by the dimensionless temperature $\Theta =k_{B}T/
E_{F,0}$ where $E_{F,0}$ is $\mu$
at $T=0$.
The Fermi-Dirac distribution function may be approximated safely by the
Maxwell-Boltzmann distribution if
\begin{equation}
\exp\{(E-\mu)/k_{B}T\} \gg 1 \mbox{    {\rm or}    }
-\mu/k_{B}T > x,\label{small}
\end{equation}
where $x\sim3-5$.
When this condition holds, one can show (by expressing the equilibrium
carrier density in terms of $E_{F,0}$) that, for
a parabolic band,
\begin{equation}
\Theta = \Bigl[ {\exp(-{\mu\over k_B T})\over {d\over 2} \Gamma({d\over
2})}\Bigr]^{2/d}.\label{CRITERION}
\end{equation}
Here $\Gamma$ is the gamma function and
$\mu$ is measured with respect to the band edge.
Eqs.  (\ref{small}) and
(\ref{CRITERION}) with $x=3$ imply that for
$\Theta < 6, 20$ and $512$
in three, two and one dimensions, respectively, the system cannot
be considered nondegenerate and $\gamma_{\rm in}$
may be significant.
Since the system dimension
dependence of the criterion
is approximately $\Theta >\exp({2x/d})$,
the condition is much more stringent
in lower dimensions, and
therefore $\gamma_{\rm in}$ can contribute significantly
even at relatively high temperatures in low-dimensional systems.

Recently, two of us\cite{HU} presented a calculation of the
finite-temperature scattering rate for hot electrons injected into
{\em n}-type GaAs which neglected the hole-scattering term.
The transition rates $P({\bf q},\omega)$ were calculated in the Born
approximation
using the total dielectric function of a
coupled electron-phonon system and the random phase approximation for
the electron polarizability.
Figure 1 shows how the result is changed when the neglected term is included
for $n=8\times10^{17}$ cm$^{-3}$ and temperatures 0 to 300 K.
At 300 K at this density, $\Theta = 0.55$ and
the system is semidegenerate.
A comparison of the
results at low energies above $E_{F,0}=47$ meV is of interest since
Heiblum, Galbi, and Weckwerth\cite{HE} experimentally determined
$\gamma$ for low energy electrons in {\em n}-type
GaAs at this doping level and low temperature (4.2 K).
Figure 1 shows that as the temperature is
increased and $f_0$ becomes nonnegligible above $\mu$,
the decay rate $\gamma$ can be significantly larger than $\gamma_{\rm out}$.

Even a relatively small enhancement of $\gamma$ due to
$\gamma_{\rm in}$ can have important experimental consequences.
For example, in ballistic electron transport and spectroscopy experiments,
one measures the fraction $\alpha$ of injected electrons that traverse
a thin transistor base without scattering.
This quantity $\alpha$ is determined by the part of
$\delta\!f$ that does not decay
during the time $t_0$ it takes for the electrons to traverse the base.
The solution of
Eq.\ (\ref{eq4b}) shows that $\alpha\propto\exp(-\gamma t_{0})$
is {\em exponentially} sensitive to the scattering rate.
Therefore, a small change in $\gamma$
can imply a substantial change in $\alpha$ or in the
differential current gain $\beta=
\alpha/(1-\alpha)$ of a hot electron transitor. The enhancement of
$\gamma$ due to $\gamma_{\rm in}$ makes devices
of this type with useful gain at room temperature less
feasible than the earlier calculation\cite{HU} indicates.

Another experimental situation where the enhancement of $\gamma$
can be significant is femtosecond optical spectroscopy,
which measures carrier thermalization due to the fastest relaxation mechanisms.
Knox and coworkers\cite{KNO} used
near-bandgap excitation and detection of carriers to study carrier-carrier
scattering in modulation-doped quantum wells. The experiments
 were performed at room temperature and doping densities
$n=3.5\times10^{11}$ cm$^{-2}$ such that
$\Theta \sim 2$. Thus the photoexcited carriers scatter from a
sea of equilibrium carriers in the semidegenerate regime and the
photoexcitation energy ($\sim$ 20 meV) is within the thermal width of the
Fermi surface, making the scattering-in contribution to the thermalization
rate significant.

Inelastic scattering also determines whether coherent or incoherent
tunneling occurs in double barrier structures.\cite{TUN}
  The scattering-in
contribution might be important when
there is a large build-up of charge in the structure at the resonant
bias voltage (as evidenced by increased photoluminescence
linewidths\cite{CHARGE}), which implies a large
$f_0({\bf k})$ in the well, where ${\bf k}$ is the momentum parallel
to the barrier planes.

Finally, we mention that for bosons, the scattering-in term has the
{\em opposite} sign from the scattering-out term since an
extra nonequilibrium occupation of a boson state
{\em enhances} rather than restricts processes which further
increase that occupation.
Thus the decay of a nonequilibrium exciton
or phonon population will be slower
(by a factor of $[1 + f_0({\bf k})]^{-1}$) than
if no scattering-in processes occurred. This is relevant to experiments on
excitonic absorption in the presence of an exciton gas
or an electron-hole pair gas.\cite{BOSE}

In conclusion, the decay of a nonequilibrium perturbation of
an electron distribution function is given by the sum of {\em both}
the equilibrium scattering-out {\em and} scattering-in rates.
We find that the scattering-in term, which is usually ignored in
scattering rate calculations, enhances the decay rate $\gamma({\bf k})$
by the factor $[1-f_0({\bf k}]^{-1}$ over $\gamma_{\rm out}({\bf k})$
and can be significant for low-energy
electrons injected into semidegenerate systems, especially in
lower dimensional systems.
Workers in the semiconductor
 field should be alerted to the possible significance of
scattering-in processes in modern device and experimental structures
which are commonly in the semidegenerate regime.
While the extremely degenerate ($k_B T/E_F \ll
1$) and nondegenerate ($k_B T/E_F \gg 1$) cases, applying
respectively to metals and lightly doped semiconductor devices, do not
involve significant scattering-in effects, heavily doped
semiconductors at room temperatures or below may be
significantly affected.

BAS thanks J.R. Lowney and P.B. Allen for helpful discussions.
Work by BY-KH and SDS is supported by the U.S. ONR and the U.S. ARO.

\vspace{7mm}
\noindent{\Large \bf Figure caption}
\vspace{5mm}

\noindent
Fig.\ 1.\ \ \
Comparison of the scattering-out rate
$\gamma_{\rm out}({\bf k})$ (thin lines), and the
decay rate of the nonequilibrium part of the distribution
function, $\gamma({\bf k}) = \gamma_{\rm out}({\bf k}) +
\gamma_{\rm in}({\bf k})$ (bold lines),
in $n$-GaAs doped at $8\times 10^{17}\,
{\rm cm}^{-3}$, for various temperatures.   The enhancement of
$\gamma({\bf k})$ over $\gamma_{\rm out}({\bf k})$
is due to the Pauli priciple,
which (for $\delta\!f({\bf k}) >0$)
restricts particles from scattering {\sl into} state $|{\bf k}\rangle$,
thus reducing the replenishment rate of $|{\bf k}\rangle$.
The cusps in the $T=0$ curve are due to plasmon and phonon emission
thresholds.\cite{HU}


\begin{references}

\bibitem[*]{NIST}Contribution of the National Institute of Standards
and Technology; not subject to copyright.

\bibitem[\dag]{FEL}NRC-NIST Postdoctoral Research Associate

\bibitem[1]{YOK} N. Yokoyama, H. Ohnishi, T. Mori, M. Takatsu, S. Muto,
K. Imamura, and A. Shibatoni in {\em Hot Carriers in Semiconductor
Nanostructures}, edited by J. Shah (Academic Press, Boston, 1992), p.443.

\bibitem[2]{BAL} A. Pavlevski, M. Heiblum, C.P. Umbach, C.M. Knoedler,
A.N. Broers, and R.H. Koch, Phys. Rev. Lett {\bf 62}, 1776 (1989).

\bibitem[3]{TUN} F. Capasso, S. Sen, F. Beltram, and A.Y. Cho in {\em Physics
of Quantum Electron Devices}, edited by F. Capasso (Springer-Verlag, Berlin,
1990), p. 181.

\bibitem[4]{LEV} A.F.J. Levi, J.R. Hayes, P.M. Platzman, and W. Wiegmann,
Phys. Rev. Lett. {\bf 55}, 2071 (1985);
R. Jalabert and S. Das Sarma, Phys. Rev. B, {\bf 41}, 3651 (1990).

\bibitem[5]{HE} M. Heiblum, D. Galbi, and M. Weckwerth, Phys. Rev. Lett.
 {\bf 62}, 1057 (1989).

\bibitem[6]{KNO} W.H. Knox, Solid State Electron. {\bf 32}, 1057 (1989).

\bibitem[7]{MANYBODY}
This agrees with the many-body theory result for the quasiparticle lifetime
(which gives the decay of the amplitude of a particle inserted
into $|{\bf k}\rangle$),
\begin{equation}
\tau ({\bf k}) = 1/\gamma({\bf k}) =
\hbar /[2|{\rm Im} \{\Sigma^{ret} ({\bf k},E({\bf k}))\}|].\nonumber
\end{equation}
$\Sigma^{\rm ret}$ is the retarded self-energy,
which can be written as
\begin{equation}
{\rm Im}[\Sigma ^{\rm ret}]= (\Sigma^{>}+\Sigma^{<})/2.\nonumber
\end{equation}
where $\Sigma^{>}/\hbar$ and $\Sigma^{<}/\hbar$ are physically
equivalent to the scattering-out and scattering-in rates,
respectively (see, {\em e.g.}, R. Jalabert and S. Das Sarma,
Phys.\ Rev.\ B, {\bf 40}, 9723 (1989), or ref. 8)

\bibitem[8]{KAD} L.P. Kadanoff and G. Baym,
 {\em Quantum Statistical Mechanics}
(Benjamin, 1962, reprinted by
Addison-Wesley, Redwood City, CA, 1989), pp. 36-40.

\bibitem[9]{ZER}
M.E. Kim, A. Das, and S.D. Senturia, Phys. Rev. B {\bf 18},
6890 (1978); P. Nozi\`{e}res
and D. Pines, Nuovo Cimento {\bf 9}, 470 (1958).

\bibitem[10]{CLA} J.F. Young, P.J. Kelly, N.L. Henry, and M.W.C.
Dharma-wardana,
Solid State Comm. {\bf 78}, 343 (1991);
J.M. Rorison and D.C. Herbert, J. Phys. C {\bf 19}, 6357 (1986).

\bibitem[11]{HU} B.Y.-K. Hu and S. Das Sarma, Phys. Rev. B {\bf 44}, 8319
(1991).

\bibitem[12]{CHARGE} M.S. Skolnick, P.E. Simmonds, D.G. Hayes, C.R.H. White,
L. Eaves, A.W. Higgs, M. Henini, O.H. Hughs, G.W. Smith, and C.R. Whitehouse,
Semicond. Sci. Technol. {\bf 7}, B401 (1992).

\bibitem[13]{BOSE}D.R. Wake, H.W. Yoon, J.P. Wolfe, and H. Morko\c{c},
Phys. Rev. B {\bf 46}, 13452 (1992).

\end{references}
\end{document}